\documentclass[multphys,vecphys]{svmult}

\usepackage{makeidx}         
\usepackage{graphicx}        
\usepackage{multicol}        
\usepackage[bottom]{footmisc}

\makeindex             

\newcommand{\beqn}{\begin{eqnarray}}
\newcommand{\eeqn}{\end{eqnarray}}
\newcommand{\be}{\begin{equation}}
\newcommand{\ee}{\end{equation}}

\def\Journal#1#2#3#4{{#1} {\bf #2}, #3 (#4)}

\def\PLB{{\em Phys. Lett.}  B}

\def\PRD{{\em Phys. Rev.} D}

\begin{document}

\title*{Upper limits on sparticle masses from WMAP dark matter 
        constraints with modular invariant soft breaking}
\author{Utpal Chattopadhyay\inst{1}\and
Pran Nath\inst{2}}
\institute{Department of Theoretical Physics, Indian Association for
the Cultivation of Sciences, Jadavpur, Kolkata 700032, India 
\texttt{tpuc@iacs.res.in}
\and Department of Physics, Northeastern University, Boston, MA, 02115, USA
 \texttt{nath@neu.edu}}
%
%
\maketitle
An analysis of dark matter within the framework of modular invariant
soft breaking is given. In such scenarios inclusion of the radiative
electroweak symmetry breaking constraint determines  
$\tan\beta$ which leads to a  more constrained analysis. 
It is shown that for $\mu$ positive for this constrained system the
WMAP data leads to upper limits on sparticle masses that lie within
reach of the LHC with also the possibility that some sparticles may
 be accessible at RUNII of the Tevatron.

\section{Introduction}
In this talk we  will focus on modular invariant soft breaking 
and an analysis of dark matter within this framework\cite{cnmodular}. We will 
then show the constraints of WMAP\cite{bennett,spergel}, the flavor changing 
neutral current
constraint arising from $b\rightarrow s+\gamma$\cite{cleo,bsgamma,bsgammanew,gambino}
 and the constraints
of radiative electroweak  symmetry breaking (REWSB) put stringent limits 
on the sparticle  masses. Specifically we will show that for the case
of $\mu>0$ the WMAP constraints lead to upper limits on sparticle masses
which all lie within the reach of the Large Hadron Collider (LHC). 
Further, it is found that some of these particles may also lie within
reach of RUNII of the Tevatron. An analysis of dark matter detection
rates is also given and it is shown that for $\mu>0$ the WMAP data 
leads to direct detection rates which lie within reach of the current and the next
generation of dark matter 
detectors\cite{cdms,genius,cline,smith,cdmsmay2004,edelweiss,damaresult}.
 For the case of $\mu<0$ the
detection rates will be accessible to the future dark matter detectors
for  a part of the allowed parameter space of the models with modular
invariant soft breaking and consistent with WMAP and the FCNC constraints.
The outline of the rest of the paper is as follows: In Sec.2 we give a brief
discussion of modular invariant soft breaking and a determination of $\tan\beta$
with radiative electroweak symmetry breaking constraints.
 In Sec.3 we give an analysis of the satisfaction of the relic density constraints 
 consistent with WMAP and upper limits on sparticle masses for $\mu>0$. In Sec.4 
 we discuss the direct detection rates. Conclusions are given in Sec.5.

\section{Modular invariant soft breaking}
We begin with string theory motivation for considering a modular invariant low
energy theory. It is well known that in orbifold string models one has a so called large 
radius- small radius symmetry 
\beqn
R\rightarrow \alpha'/R
\label{1}
\eeqn
More generally one has an $SL(2,Z)$ symmetry and such a symmetry is valid even 
non-perturbatively which makes it very compelling that this symmetry survives in the 
low energy theory. 
In formulating an effective low energy theory it is important to simulate as
much of the symmetry of the underlying string theory as possible.
This provides the motivation for considering low energy 
effective theories with modular 
invariance\cite{fmtv,brignole,nilles,gaillard}..
 With this in  mind we consider
an effective four dimensional theory arising from string 
theory assumed to have a target space modular $SL(2,Z)$ invariance 
  \beqn
  T_i\rightarrow
T'_i=\frac{{a_iT_i-ib_i}}{{ic_iT_i+d_i}},\nonumber\\
\bar T_i\rightarrow
\bar T'_i=\frac{a_i\bar T_i+ib_i}{-ic_i\bar T_i+d_i},\nonumber\\
(a_id_i-b_ic_i)=1,~~~ (a_i,b_i,c_i,d_i \in Z).
\label{2}
\eeqn
Under the above transformation  the superpotential and the K\"ahler potential transform 
but the combination
\beqn
G=K+ln(WW^{\dagger})
\eeqn
is invariant. Further, the  scalar potential $V$ defined by

$$ V= e^{G}((G^{-1})^i_jG_iG^j+3) +V_D $$
is also invariant under modular transformations. We require that $V_{soft}$ also maintain 
modular invariance and indeed this invariance will naturally be maintained in our analysis.
Typically chiral fields, i.e., quark, leptons and Higgs fields will transform under 
modular transformations and for book keeping it is useful to assign modular weights to
operators. 
Thus a function $f(T_i,\bar T_i)$ has modular weights $(n_1,n_2)$ if 
  \beqn
f(T_i,\bar T_i) \rightarrow
(icT_i+d)^{n_1} (-ic\bar T_i+d)^{n_2} f(T_i,\bar T_i) 
\eeqn
Below we give a list of modular weights for a few cases.

\begin{table}[hbt]
{\centering 
\begin{tabular}{|c|c|}
\hline
quantity   &  modular weights $(n_1,n_2)$\\
\hline
$|W|$                          &  $(-\frac{1}{2}, -\frac{1}{2})$    \\
\hline
$e^{i\theta_W}$     &   $(-\frac{1}{2}, \frac{1}{2})$       \\
\hline
 $\eta(T_i)$   &      $(\frac{1}{2}, 0)$                \\
\hline
 $2\partial_{T_i}ln\eta(T_i)+(T_i+\bar T_i)^{-1}$  & $(2,0)$    \\
\hline
  $\partial_{T_i} W - (T_i+\bar T_i)^{-1}W$    & $(1,0)$   \\
  \hline
 $(T_i+\bar T_i)$ & $(-1,-1)$ \\
 \hline
  $|\gamma_s|$  & $(0,0)$ \\
\hline
  $|\gamma_{T_i}|$       &  $(0,0)$                   \\
\hline
   $e^{i\theta_{T_i}}$   &   $(1,-1)$    \\
\hline
 $e^{i\theta_{S}}$   &   $(0,0)$    \\
\hline
$A^{0}_{\alpha\beta\gamma}$ & $(1,0)$ \\
\hline
$B^{0}_{\alpha\beta}$ & $(1,0)$ \\
\hline
$1/\sqrt{f} =1/(\prod (T_i+\bar T_i))^{\frac{1}{2}}$  &  ($\frac{1}{2}$,$\frac{1}{2}$) \\
\hline                      
\hline
\end{tabular}
\par}
\centering
\caption{A list of modular weights under the modular 
transformations.}
\end{table}

%
\begin{figure}
\centering
\includegraphics[height=4cm]{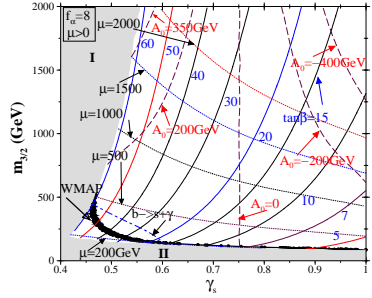}
%
%
\caption{Plot is given of the  
contours of constant  $A_0, \mu, \tan\beta$ in the
 $(\gamma_s-m_{3/2})$ plane for the case $\mu>0$. 
The constraint of $b \rightarrow s + \gamma$ decay  
is shown as a dot-dashed line below which the region is 
disallowed.   
The region where the WMAP relic density constraint is satisfied is shown 
 as small 
shaded area in black. The gray region-I refers to the  discarded 
region  with large $\tan\beta$ where 
Yukawa couplings lie beyond the perturbative domain.
The gray region II arises from 
the absence of REWSB or a  $m_{\tilde \chi_1^\pm}$ below 
the experimental limit.  Taken from Ref.\cite{cnmodular}.}
\label{fig1}       
\end{figure}
%

%
\begin{figure}
\centering
\includegraphics[height=4cm]{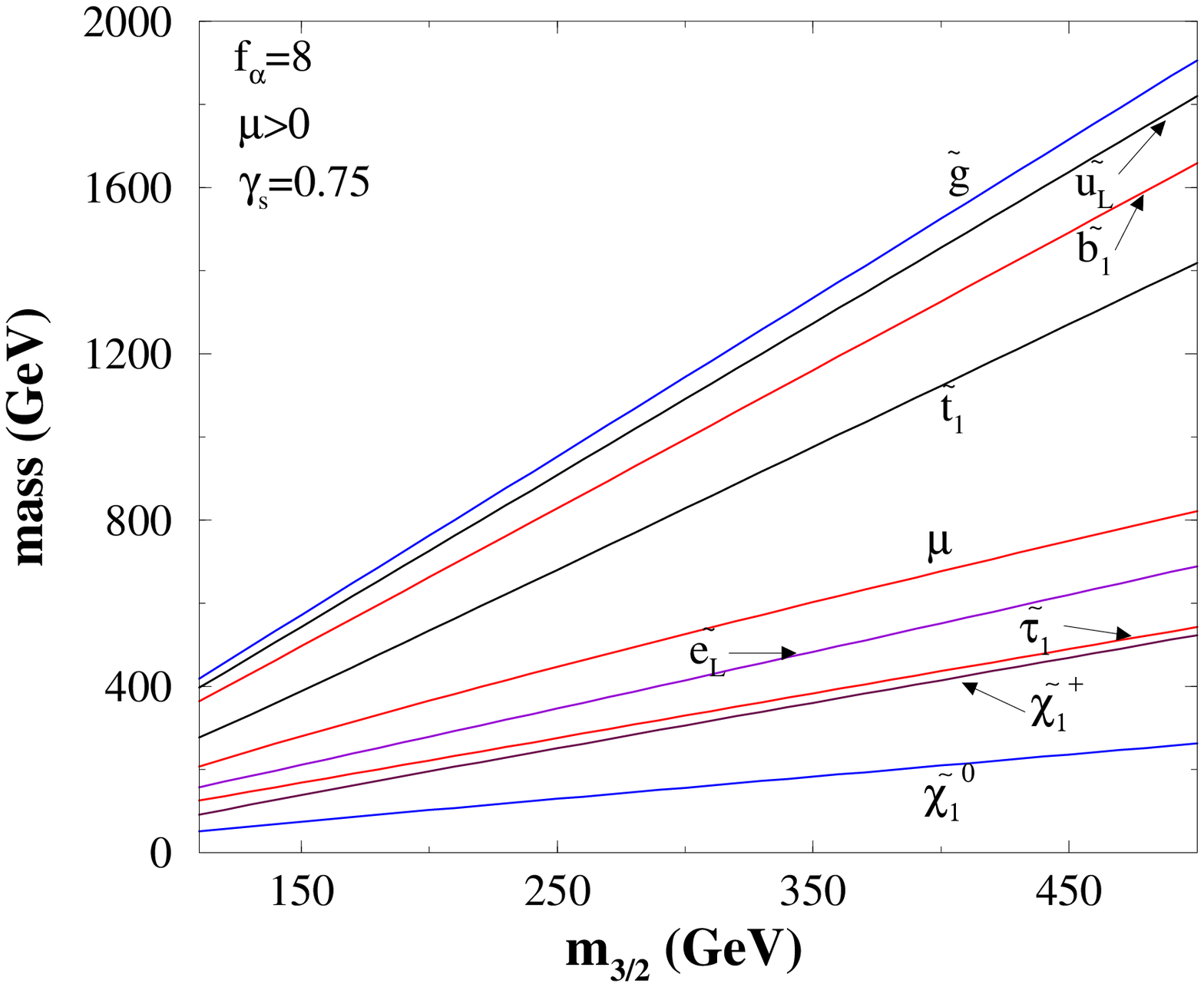}
%
%
\caption{An exhibition of the 
variation of sparticle masses with  
$m_{3/2}$ with $\gamma_s=0.75$ for the case when $\mu>0$.
The WMAP constraint is not exhibited. Taken from Ref.\cite{cnmodular}}
\label{fig2}       
\end{figure}
%

\subsection{Modular invariant $V_{soft}$} 
We begin by considering the condition for the vanishing of the 
vacuum energy. Using the supergravity form of the scalar potential
the condition that vacuum energy vanish is given by 
\beqn
 |\gamma_S|^2+\sum_{i=1}^3 |\gamma_{T_i}|^2 =1 
 \eeqn
 where we have defined $\gamma_s$ and $\gamma_{T_i}$ as follows
\beqn 
\gamma_s= (S+\bar S)G,S/\sqrt 3 
= |\gamma_S| e^{i\theta_S}
\eeqn
\beqn
 \gamma_{T_i}= (T_i+\bar T_i)G,{T_i}/\sqrt 3
 = |\gamma_{T_i}| e^{i\theta_{T_i}}
 \eeqn
 In the investigation of soft breaking we follow the usual procedure
 of supergravity where one has a visible sector and a hidden sector 
 and supersymmetry breaking occurs in the hidden sector and is 
 communicated to the visible sector by gravitational interactions.
 For the analysis here we choose the hidden sector to be of the
 form\cite{nathtaylor} 
  \beqn
  W_h=F(S)/\prod \eta(T_i)^2
  \eeqn
and for the Kahler potential we choose 
\beqn
K=D(S,\bar S) -\sum_iln(T_i+\bar T_i) +
\sum_{i\alpha}(T_i+\bar T_i)^{n_{\alpha}^i}
C_{\alpha}^{\dagger}C_{\alpha}
\eeqn
where $C_{\alpha}$ are the chiral fields.
Using the technique of supergravity models\cite{can} the soft breaking potential $V_{soft}$  
is given by\cite{nathtaylor}(for previous analyses see Refs.\cite{brignole,gaillard,kane} 
\beqn
V_{soft}= m_{3/2}^2 \sum_{\alpha}(1+3\sum_{i=1}^3n^i_{\alpha}|\gamma_{T_i}|^2)
c_{\alpha}^{\dagger}c_{\alpha}+ (\sum_{\alpha\beta}B^0_{\alpha\beta}w^{(2)}_{\alpha\beta} + 
\sum_{\alpha\beta\gamma} 
A_{\alpha\beta\gamma}^0w_{\alpha\beta\gamma}^{(3)}+ H.c.)
\eeqn
where
\beqn
w^{(2)}_{\alpha\beta} = \mu_{\alpha\beta} C_{\alpha}C_{\beta}\nonumber\\  
w^{(3)}_{\alpha\beta\gamma} = Y_{\alpha\beta\gamma}
C_{\alpha}C_{\beta}C_{\gamma}
\eeqn
The soft breaking parameters $A^0$ and $B^0$ may be expressed in the form
\beqn 
A_{\alpha\beta\gamma}^0= -\sqrt 3 m_{3/2} \frac{e^{D/2-i\theta_W}}{\sqrt f}  
[ |\gamma_{S}|  e^{-i \theta_S} 
 (1-(S+\bar S)\partial_S lnY_{\alpha\beta\gamma})\nonumber\\ 
 +\sum_{i=1}^3 |\gamma_{T_i}| 
e^{-i \theta_{T_i}}(1+n^i_{\alpha}+n^i_{\beta} +n^i_{\gamma} 
 -(T_i+\bar T_i)\partial_{T_i}ln Y_{\alpha\beta\gamma}
 -(T_i+\bar T_i) n^i_{\alpha\beta\gamma} G_2(T_i))] 
\nonumber
\eeqn

\beqn 
B_{\alpha\beta}^0= -m_{3/2} \frac{e^{D/2-i\theta_W}}{\sqrt f}  
[ 1+ \sqrt 3  |\gamma_{S}|  e^{-i \theta_S} 
 (1-(S+\bar S)\partial_S ln\mu_{\alpha\beta})\nonumber\\ 
 + \sqrt 3\sum_{i=1}^3 |\gamma_{T_i}| 
e^{-i \theta_{T_i}}(1+n^i_{\alpha}+n^i_{\beta} 
 -(T_i+\bar T_i)\partial_{T_i}ln\mu_{\alpha\beta}
 -(T_i+\bar T_i) n^i_{\alpha\beta} G_2(T_i)
 )] 
\nonumber
\eeqn
and further the universal gaugino mass is given by
\beqn 
m_{1/2} = \sqrt 3 m_{3/2} |\gamma_s| e^{-i\theta_S}  
\eeqn

\subsection{Determination of $\tan\beta$ from modular invariant soft
breaking and EWSB constraints} 
We begin with  a discussion of the front factor that appears in $A^0$ and
$B^0$\footnote{This front factor is quite general and also appears in soft breaking arising from the
intersecting D brane models\cite{kn1}.}
\beqn
Front ~factor  = e^{D/2-i\theta_W}/\sqrt f
\eeqn
The front factor has a non vanishing modular weight and 
 the modular invariance of $V_{soft}$ cannot be maintained without it. 
There are two main elements in this front factor which are of interest to us
here. First, there is factor of of $1/\sqrt f$ or a factor
\beqn
1/\sqrt {\prod (T_i+\bar T_i)} 
\eeqn
which produces several solutions to the soft parameters 
at the self dual points $T_i= (1, e^{i\pi/6})$ so that
\beqn
f= 8, 4\sqrt 3, 6, 3\sqrt 3
\eeqn
If we  include the complex structure moduli $U_i$ then
\beqn
 \prod (T_i+\bar T_i)\rightarrow \prod (T_i+\bar T_i) (U_i+\bar U_i)\nonumber\\
f=2^n3^{3-\frac{n}{2}}~~ (n=0,..,6)
\eeqn
Assuming that the minimization of the potential occurs at one of these
self dual points one finds that there is a multiplicity of soft parameters 
all consistent with modular invariance. Of course, it may happen
that the minimization occurs away from the self dual points. In this case 
there the f factor will take values outside of the sets given above. 
The second element that is of interest to us in the front factor is the 
quantity  $e^{D/2}$. This factor is of significance since it can be related 
 to the string gauge coupling constant $g_{string}$ so that
 \beqn
 e^{-D}=\frac{2}{g_{string}^2}
 \eeqn
The importance of front factor becomes clear when one considers the 
electroweak symmetry breaking constraints arising from the minimization
of the potential with respect to the Higgs vacuum expectation values 
$<H_1>$ and $<H_2>$. In supergravity models one of these relations is used
to determine $\mu$ and the other relates the soft  parameter $B$ to $\tan\beta$.
In supergravity one uses the second relation to eliminate $B$ in favor of $\tan\beta$.
However, in the model under consideration $B$ is now determined and thus
the second minimization constraint allows one to determine  $\tan\beta$ in terms
of the other soft parameters and $\alpha_{string}= g_{string}^2/4\pi$. 
Thus specifically the second  constraint reads 
\beqn
 -2\mu B= \sin 2\beta (m_{H_1}^2+m_{H_2}^2 +2\mu^2)
\eeqn
Turning this condition around we determine $\tan\beta$ such that 
   \beqn
\tan\beta=
\frac{(\mu^2+ \frac{1}{2} M_Z^2 +m_{H_1}^2) f_{\alpha}^{1/2} } 
{\sqrt{2\pi}\mu m_{3/2} \tilde r_B\alpha_{string} }\nonumber\\
(|-1+3\sum_i |\gamma_i|^2-\sqrt 3|\gamma_S|
(1-(S+\bar S)\partial_S ln\mu)|)^{-1}
\label{tan}
\eeqn 
There is one subtle point involved in the implementation of this equation.
One is a relation that holds at the tree level and is accurate only at 
scales where the one loop correction to this relation is small. This
happens when $Q\sim m_{\tilde t}$ or $Q\sim$ (highest mass of the spectrum)/2.
Thus for the relation of Eq.(\ref{tan}) to be accurate we should use the
renormalization group improved values of all the quantities on the right hand
side of Eq.(\ref{tan}). This is specifically the case for the Higgs mass parameters
and $\mu$. One obtains their values at the high scale $Q$ by running the renormalization
group equations between $M_Z$ and $Q$.  
The general analysis used is that of renormalization group analysis of
supergravity theories (see, e.g., Ref.\cite{an}).
Determination of $\tan\beta$ is done in an iterative procedure.
One starts with an assumed value of $\tan\beta$ and then one determines
$\mu$ through radiative breaking of the electroweak symmetry,  one 
determines the sparticle masses and the Higgs masses and uses these in 
Eq.(\ref{tan}) to determine the new value of $\tan\beta$. This iteration
continues till  consistency is obtained. 
Quite interestingly there are solutions to the iterative procedure,
and the convergence is quite rapid. Thus  $\tan\beta$ is uniquely 
determined for each point in the space of other soft parameters
provided radiative electroweak symmetry breaking constraints are satisfied.
In the analysis the Higgs mixing parameter $\mu$  and specifically its sign
plays an important role. Interestingly there is important correlation between
the sign of the supersymmetric contribution to the anomalous magnetic moment of 
the muon\cite{yuan} and the sign of the $\mu$ parameter.  
It turns out the current data seems to indicate a positive supersymmetric 
contribution and a positive $\mu$\cite{chattog2}. Thus in the analysis we will
mainly focus on $\mu$ positive. However, for the sake of completeness we will
also include in our analysis the $\mu<0$ case.

%
\begin{figure}
\centering
\includegraphics[height=4cm]{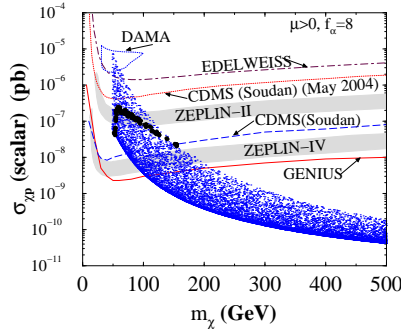}
%
%
\caption{A scatter plot of the spin  
independent LSP-proton cross section vs LSP mass for the case  
$\mu>0$ when $\gamma_s$ and $m_{3/2}$ are integrated. 
The region with black circles satisfies the WMAP constraint. Present 
limits (top three contours) and 
future accessibility regions are shown.
Taken from Ref.\cite{cnmodular}}
\label{fig3}       
\end{figure}
%

%
\begin{figure}
\centering
\includegraphics[height=4cm]{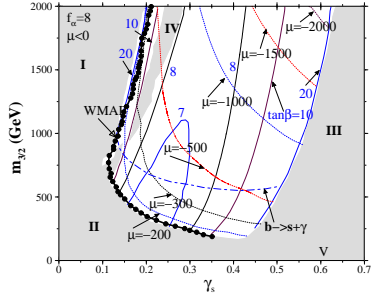}
%
%
\caption{
Plot is given of the contours of constant $\tan\beta$ and $\mu$ 
in the $(\gamma_s-m_{3/2})$ plane for the case $\mu<0$.
The constraint of $b \rightarrow s + \gamma$ decay  
is shown as a dot-dashed line below which the region is 
disallowed. The region where the WMAP relic density constraint is satisfied 
is shown  as small shaded area in black.
The gray region I and III are
disallowed because of the absence of consistent GUT scale
inputs. The region II refers to absence of REWSB or smaller
than experimental lower limits of $m_{\tilde \chi_1^\pm}$.
The region IV is a no solution zone like I and
III, but its location and extent depends on the sensitivity of
the minimization scale for REWSB. Region V is the tachyonic
$\tilde \tau_1$ zone.
Taken from Ref.\cite{cnmodular}}
\label{fig4}       
\end{figure}

\section{Analysis of supersymmetric dark matter}
There is already a great deal of analysis of supersymmetric dark matter
in the literature (For a sample of recent analyses\cite{related} 
see Refs.\cite{ccn2,Chattopadhyay:2003yk,Baer:2003ru,Binetruy:2003yf,adkt,rosz})..
Specifically, over the past year analyses of dark
matter matter have focussed on including the constraints of 
WMAP\cite{Chattopadhyay:2003xi,elliswmap,hb/fp,Chattopadhyay:2003qh,gomez}
Here we discuss the analysis of dark matter within the framework of modular 
invariant soft breaking where $\tan\beta$ is a determined quantity. 
Thus using the 
sparticle spectra generated by the procedure of Sec.2 one can
compute the relic density of lightest neutralinos within the modular invariant framework. 
Quite interesting is the fact that the relic 
density constraints arising from WMAP data are satisfied by the
modular invariant theory in the  determined $\tan\beta$ scenario. 
It is also possible to satisfy the FCNC constraints. One finds that
the simultaneous imposition of the WMAP relic density constraints and
of the FCNC constraints leads to upper limits on the sparticle masses
for the case of $\mu$ postive. The sparticle spectrum that is 
predicted in this case can be fully tested at the LHC.
Further, a  part of the parameter space is also accessible at the Tevatron. 

We discuss the results now in a quantitative fashion. In Fig.(1) a plot is given
of the contours of constant $A_0$, constant $\mu$ and constant $\tan\beta$ in the
$m_{3/2}-\gamma_S$ plane. One finds that there are regions where the relic density
constraints consistent with the WMAP data and the FCNC constraints are satisfied.
The value of $m_{3/2}$ consistent with all the constraints has an upper limit of 
about 350 GeV. In Fig.(2) a plot of the sparticle spectrum as a function of $m_{3/2}$
is given for $\gamma_S=0.75$. One finds that the sparticle masses with 
$m_{3/2}<350$ GeV lie in a range accessible at the LHC. In fact, for a range of the 
parameter space some  of the sparticles may also be accessible at the Tevatron.
Thus much of the Hyperbolic Branch/Focus Point (HB/FP) region\cite{Chan:1997bi}
 seems to be eliminated
by the constraints of WMAP and FCNC within the modular invariant soft breaking\cite{cnmodular}.

In Fig.(3) an analysis of the direct detection cross-section for $\sigma_{\chi-p}$ 
as a function of the LSP mass is given. One finds that all of the parameter space
of the model will  be probed in the current and future dark matter colliders. 
An analysis analogous to that of Fig.(1) but for $\mu<0$ is given in Fig.(4) while
 an analysis analogous to Fig.(3) is given in Fig.(5). In this case one finds that
 a part of the parameter space consistent with WMAP can be probed in the current and
future dark matter experiments. Finally, the analysis presented above is done under the
assumption that the chiral fields have zero modular weights. For non-vanishing modular
weights one needs a realistic string model and an analysis of the sparticle spectra and
dark matter for such a model  should be worthwhile using the above framework.

%
\begin{figure}
\centering
\includegraphics[height=4cm]{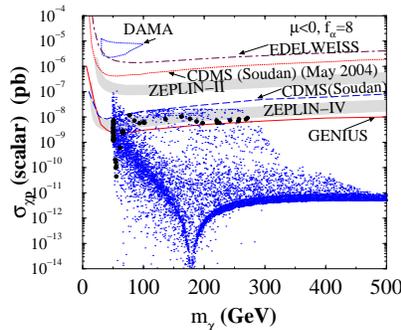}
%
%
\caption{A scatter plot of the 
spin-dependent LSP-proton cross section vs LSP mass for the case when $\mu<0$. 
The region with black circles satisfies the WMAP constraint.
Taken from Ref.\cite{cnmodular}.}
\label{fig5}       
\end{figure}

\section{Conclusion}
In this paper we have analyzed the implications of modular invariant soft breaking in a 
generic heterotic string scenario under the constraint of radiative breaking of the 
electroweak symmetry. It was shown that in models of this
type $\tan\beta$ is no longer an arbitrary parameter but a determined quantity. 
Thus the constraints of modular invariance along with a determined $\tan\beta$ reduced
the allowed parameter space of the model.
Quite remarkably  one finds that the reduced  parameter space 
allows for the satisfaction of the accurate relic  density constraints
given by WMAP. Further, our analysis shows that the WMAP constraint combined 
with the FCNC constraint puts upper limits on the sparticle masses for the
case $\mu>0$ which are remarkably low implying that essentially all of the 
sparticles would be accessible at the LHC and some of the sparticles may also be
visible at the Tevatron. Further, we analysed the direct detection rates in dark 
matter detectors in such a scenario. It is found that for the case $\mu>0$ 
the dark matter detection rates fall within the sensitivities of the current and
future dark matter detectors. For the case $\mu<0$ a part of the allowed parameter
space will be accessible to dark matter detectors. It should be of interest to analyze
scenarios of the type discussed above with determined $\tan\beta$ in the
investigation of other SUSY phenomena. Further, it would be interesting to examine if 
similar limits arise in models with modular invariance in extended MSSM seenarios, such
as the recently proposed Stueckelberg extension of MSSM\cite{kn2}.

\noindent
{\bf Acknowledgements}\\
This work is supported in part by NSF grant PHY-0139967.

\end{document}